\documentclass[aps,prl,twocolumn,superscriptaddress,floatfix]{revtex4}
\usepackage[utf8]{inputenc}
\usepackage{tabularx}
\usepackage{array}
\usepackage{float}
\usepackage{graphicx}
\usepackage{longtable}
\usepackage{amsmath}
\usepackage{amssymb}
\usepackage{textcomp}
\usepackage{color}
\usepackage{gensymb}

\newcommand{\modif}[1]{\textcolor{black}{#1}}
\newcommand{\modiff}[1]{\textcolor{blue}{#1}}

\begin{document}

\title{Robust Antiferromagnetism in Y$_{2}$Co$_{3}$}
\author{Yunshu Shi}
\affiliation{Department of Physics and Astronomy, University of California, Davis, California 95616, USA}
\author{David S. Parker}
\affiliation{Materials Science and Technology Division, Oak Ridge National Laboratory, Oak Ridge, Tennessee 37831, USA}
\author{Eun Sang Choi}
\affiliation{National High Magnetic Field Laboratory, Tallahassee, Florida 32310, USA}
\author{Kasey P. Devlin}
\affiliation{Department of Chemistry, University of California, Davis, California 95616, USA}
\author{Li Yin}
\affiliation{Materials Science and Technology Division, Oak Ridge National Laboratory, Oak Ridge, Tennessee 37831, USA}
\author{Jingtai Zhao}
\affiliation{Department of Physics and Astronomy, University of California, Davis, California 95616, USA}
\affiliation{School of Material Science and Engineering, Guilin University of Electronic Technology, Guilin, Guangxi 541004, China}
\author{Peter Klavins}
\affiliation{Department of Physics and Astronomy, University of California, Davis, California 95616, USA}
\author{Susan M. Kauzlarich}
\affiliation{Department of Chemistry, University of California, Davis, California 95616, USA}
\author{Valentin Taufour}
\email{vtaufour@ucdavis.edu}
\affiliation{Department of Physics and Astronomy, University of California, Davis, California 95616, USA}

\begin{abstract}
We report on a solution-growth based method to synthesise single crystals of Y$_{2}$Co$_{3}$ and on its structural and magnetic properties. We find that Y$_{2}$Co$_{3}$ crystallizes in the La$_{2}$Ni$_{3}$-type orthorhombic structure with space group \textit{Cmce} (No.64), \modif{with Co forming distorted Kagome lattices}. Y$_{2}$Co$_{3}$ orders antiferromagnetically below $T_\textrm{N} = 252$\,K. Magnetization measurements reveal that the moments are primarily aligned along the $b$ axis with evidence for some canting. Band-structure calculations indicate that ferromagnetic and antiferromagnetic orders are nearly degenerate, \modif{at odds with experimental results.} Magnetization measurements under pressure up to 1\,GPa reveal that the N\'{e}el temperature decreases with the slope of $-1.69$\,K/GPa. \modif{We observe a field-induced spin-flop transition in the magnetization measurements at 1.5\,K and 21\,T with magnetic field along the $b$ direction. The magnetization is not saturated up to 35\,T, indicating that the antiferromagnetic ordering in Y$_2$Co$_3$ is quite robust}, which is surprising for such a Co-rich intermetallic.
\end{abstract}

\maketitle

\section{Introduction}

Magnetic materials play an important role in the development of new energy and quantum information technologies. When driven toward an instability, they can also show novel physical properties such as unconventional superconductivity, and challenge our theoretical understanding of quantum phenomena. Identifying materials near magnetic instabilities remains a research frontier because of the multitude of competing interactions, resulting in coupled magnetic, electronic and structural effects. Materials with unconventional behavior can sometimes be identified by looking at trends in properties that are transgressed by just a few exceptional compounds. Here we report on the physical properties of Y$_2$Co$_3$, which displays an antiferromagnetic order, despite the large Co content of the material, and the relative stability of a ferromagnetic ground state in electronic structure calculations. We also find that the antiferromagnetic order is robust against an applied magnetic field up to \modif{$35$\,T}. These results indicate that Y$_2$Co$_3$ has rather unusual magnetic properties,\modif{which is likely related to the distorted Kagome lattice of Co.}

\begin{figure}[!htb]
	\centering
	\includegraphics[width=\linewidth]{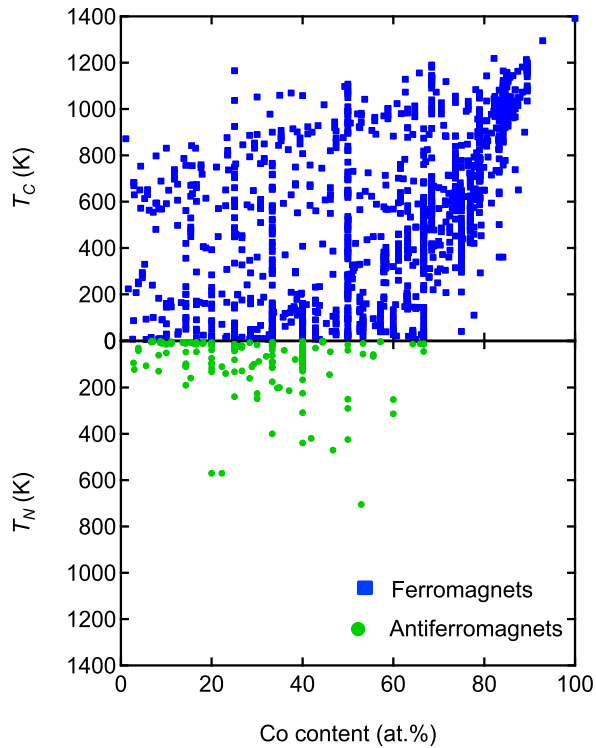} 
	\caption{Curie and N\'{e}el temperatures as a function of Co content for Co-based magnetic materials as a results of a systematic literature survey~\cite{Byland}.\label{Co compound}}  
\end{figure}

Most of the Co-based compounds are ferromagnetic, particularly with high Co content ($\geqslant$ 60\%).  
Figure~\ref{Co compound} shows the Curie and N\'{e}el temperatures as a function of cobalt content for 1511 Co-based compounds with magnetic ordering. We can see that there are no antiferromagnets among Co-based compounds with cobalt content larger than 70\,at.\%. 
\modif{With Co content no less than 50 at.\%, there are only six antiferromagnets having relatively high Néel temperature (> 100 K). These compounds contain electronegative ions (CoO, Co$_9$S$_8$~\cite{Hebert2021Transport}), or other magnetic elements (CoMn~\cite{menshikov1985magnetic}), or are well understood with band structure calculations (Ti$_2$Co$_3$Si~\cite{Hamm2016}). However, this is not the case for Y$_2$Co$_3$ and La$_2$Co$_3$.}
As a rare-earth and cobalt based compound without any electronegative anions, Y$_{2}$Co$_{3}$ doesn't seem like a candidate for antiferromagnetic ordering. We performed first principle calculations to investigate the possible magnetic structures of Y$_{2}$Co$_{3}$ and to explain its antiferromagnetic behavior. The spin-polarized density functional theory (DFT) calculations reveal that a complex antiferromagnetic state is energetically degenerate with a ferromagnetic solution, indicating the proximity to a ferromagnetic instability. However, our experimental results show that the antiferromagnetic ordering in Y$_{2}$Co$_{3}$ is quite robust and difficult to suppress with the application of relatively high magnetic fields. \modif{This illustrates that complex physics, not easily capturable from first principle calculations, can occur even in the complete absence of the 4f electrons commonly believed necessary for such behavior.}


Single crystals of Y$_{2}$Co$_{3}$ are difficult to synthesize due to yttrium reacting with common crucible materials, such as alumina, and the narrow growth region~\cite{wu1991re}. As a result, previous studies on Y$_{2}$Co$_{3}$ were limited to poly-crystalline samples and produced incomplete results regarding its crystal structure. A cubic crystal structure was reported in 1965~\cite{pelleg1965yttrium}, and an unparameterized orthorhombic crystal structure based on polycrystal studies was reported in 1992~\cite{white1992magnetic}.

Using a solution growth method in tantalum crucibles, we were able to overcome the difficulties in synthesis and produce high quality single crystals of Y$_{2}$Co$_{3}$. This enables us to identify the crystal structure as the La$_{2}$Ni$_{3}$-type which is exceptionally rare in magnetic materials: of all the rare-earth cobalt (R-Co) intermetallic compounds, only neodymium~\cite{ray1973revised}, lanthanum~\cite{Gignoux1985PB} and yttrium are known to form a stable La$_{2}$Ni$_{3}$-type structure in combination with cobalt. In addition, a few other rare earth elements (Pr, Sm, Gd,) can form such a structure, albeit with Si substitutions \cite{Tence2014Stabilization,Mahon2018JAC,Morozkin2015JSSC}.
Among all of these La$_{2}$Ni$_{3}$-type R$_{2}$Co$_{3-x}$Si$_{x}$ ($0\leq x<0.5$) compounds, only La$_{2}$Co$_{3}$ and Y$_{2}$Co$_{3}$ have an antiferromagnetic ordering. 
Other compounds in this family show ferrimagnetic orderings with Curie temperatures varying from 64\,K (Pr) to 388\,K (Gd)~\cite{Mahon2018JAC}. Interesting magnetic properties due to their complex magnetic structures were observed in this family, including strong magnetocaloric effect (Gd$_{2}$Co$_{3-x}$Si$_{x}$) and metamagnetic transition (Pr$_{2}$Co$_{2.8}$Si$_{0.2}$)~\cite{Mahon2018JAC}. However, the magnetic structures of these R$_{2}$Co$_{3-x}$Si$_{x}$ compounds are complicated by the presence of two magnetic elements: rare-earth element and Co. In Y$_{2}$Co$_{3}$, the magnetic moment is only provided by cobalt and simplifies the magnetic structure determination. Thus, in addition to investigating the origin of the robust and unexpected antiferromagnetic ordering in Y$_{2}$Co$_{3}$, understanding the magnetic structure of Y$_{2}$Co$_{3}$ will help reveal the underlying contributions from Co in these R$_{2}$Co$_{3-x}$Si$_{x}$ systems. 

\section{Methods}

\subsection{Single Crystal Growth}
The single-crystalline Y$_{2}$Co$_{3}$ samples were prepared by solution growth method~\cite{canfield2001high}. Based on the reported Y-Co binary phase diagram~\cite{wu1991re}, a starting composition of Y$_{55.5}$Co$_{44.5}$ was arc-melted and sealed in a clean tantalum crucible with a tantalum filter~\cite{Jesche2014PM}. The tantalum assembly was sealed in a silica tube with partial argon pressure. An initial temperature profile with a decantation at 820\,$\celsius$ revealed that no crystals grow above that temperature. A follow-up decantation at 760\,$\celsius$ produced large amount of YCo single crystals but no Y$_2$Co$_3$ crystals. Based on these attempts, we concluded that the previously reported binary phase diagram is inaccurate and adjusted our initial composition to Y$_{51.5}$Co$_{48.5}$. Following the same experimental method, the sample was heated up to 1150\,$\celsius$ within 4 hours and held for 5\,hours, quickly cooled down to 945\,$\celsius$ and slowly cooled down to 825\,$\celsius$ within 133\,hours. According to the previously reported Y-Co binary phase diagram~\cite{wu1991re}, a large amount of single crystal YCo$_{2}$ should also have been grown with the starting composition and temperature profile described above. However, a large amount of Y$_{2}$Co$_{3}$ single crystals with a small amount of polycrystals YCo$_{2}$ were observed, further confirming that the Y$_{2}$Co$_{3}$ part of the compositional binary phase diagram might be inaccurate.
\begin{figure}[b]
	\centering
	\includegraphics[width=\linewidth]{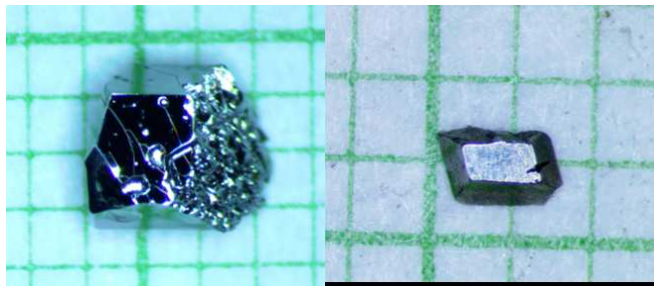} 
	\caption{Pictures of Y$_{2}$Co$_{3}$ single crystals.\label{picture}}  
\end{figure}

\subsection{Crystal Structure Identification}
Single-crystal X-ray diffraction (XRD) data were collected from the silver reflective crystal shard with a size of $0.147$\,mm $\times 0.131$\,mm $\times 0.088$\,mm at $290$\,K using a Bruker Apex-II Dual source Cu/Mo diffractometer with CCD detector, Mo(K$\alpha$) radiation, and a graphite monochromator. Only the highest-symmetry Bravais lattice suggested for the refined unit cell parameters was selected for the collected data. The frames were integrated by using SAINT program within APEX III version 2017.3-0. The centrosymmetric space group \textit{Cmce} (No.~64) was suggested by XPREP based on the analysis of systematic absences and its figure of merit. The structure was determined using direct methods, and difference Fourier synthesis was used to assign the remaining atoms (SHELXTL version 6.14)~\cite{sheldrick2008short}. \modif{Powder X-ray diffraction (PXRD) measurements were performed by the Rigaku MiniFlex600 diffractometer. }

\subsection{Crystal Orientation}
\begin{figure}[ht]
	\centering
	\includegraphics[width=\linewidth]{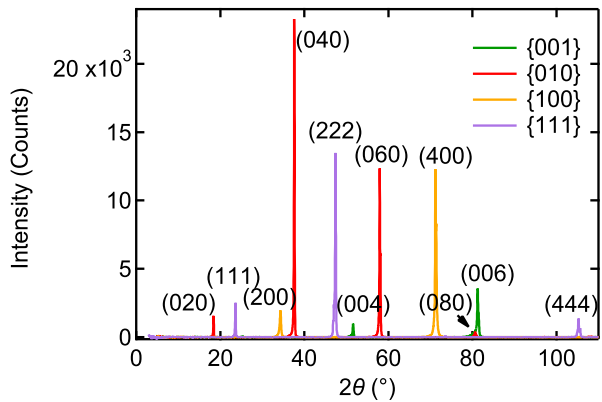}
	\caption{X-ray diffraction patterns for single crystals of Y$_2$Co$_3$ polished to show flat surfaces perpendicular to the [001], [010], [100] directions.\label{orientation}}  
\end{figure}

Two crystals are shown in Figure~\ref{picture}. For the crystal on the left, the naturally grown largest surface is the (010) plane, whereas for the crystal on the right, it is the (111) plane. This illustrate that the crystallographic orientation cannot be easily identified from the morphology of the crystals.
In order to study the anisotropic magnetic behavior of Y$_{2}$Co$_{3}$, the orientation of the single crystal was investigated with X-ray diffraction (Rigaku MiniFlex600 diffractometer) on facets~\cite{jesche2016x}. A crystal was polished in order to remove the thin layer of flux on the surface and to create parallel facets. The crystal was then placed on the puck with one facet facing upward for the XRD $2\theta$-scan. With the group of diffraction peaks, the facets were identified. For all the bulk single-crystalline Y$_{2}$Co$_{3}$ samples, only \{111\} and \{010\} plane families are naturally grown, while \{001\} and \{100\} plane families were obtained from polishing and were confirmed with the XRD result shown in Figure~\ref{orientation}.

\subsection{Physical Property Measurements}
Magnetization measurements were performed with a Magnetic Property Measurement System (MPMS, Quantum Design) in the temperature range of 2\,K to 300\,K with the applied magnetic fields up to 7\,T. Magnetization as a function of temperature was measured with an applied magnetic field of 2\,T. Magnetization vs magnetic field curves were measured at 5\,K, 100\,K and 300\,K.

\modif{The high field DC magnetization data were obtained at Cell 8 using a vibrating sample magnetometer (VSM) and a water-cooled Francis-Bitter magnet in Tallahassee, FL. A single crystal (35.7 mg) sample was attached on a polycarbonate sample holder using GE-7031 varnish for the VSM measurement. The AC susceptibility measurement was performed using a home-made AC susceptometer on the same sample measured with the VSM. The sample was cut to a smaller piece (21.9 mg) and mounted inside a Kapton tubing. The AC excitation field was fixed at 3.2 Oe and 131 Hz.}

Magnetization under pressure was measured up to 1\,GPa. Pressure was applied at room temperature using a hybrid CuBe cylindrical high pressure cell (HMD01-001-00 hydrostatic pressure cell, CC-Spr-$\Phi$ 8.5D-MC4  1.3\,GPa model) with Daphne 7373 as the pressure transmitting medium. To obtain the pressure near room temperature ($p_\textrm{290\,K}$), we calibrated the pressure by measuring the Curie temperature of gadolinium~\cite{Bartholin1968} and the superconducting transition temperature of lead~\cite{Bireckoven1988JPESI} ($p_\textrm{7\,K}$). For our measurements, we used a piece of lead next to our sample. The pressure values in this article are given at $290$\,K, using the calibrated formula ($p_\textrm{290\,K} = 0.3092 + 1.0933 p_\textrm{7\,K}$).

Resistivity and heat capacity measurements were carried out with a Physical Property Measurement System (PPMS, Quantum Design).  Resistivity data was measured using the four probe method with the current along the \textit{b} axis.

\subsection{First Principles Calculations}

In order to develop a description of the apparent complex magnetic character of Y$_2$Co$_3$, we have performed spin-polarized density functional theory calculations of this material, using the linearized augmented plane-wave code WIEN2K \cite{blaha2001}. We have used the generalized gradient approximation \cite{perdew1996}, using the experimental structure with non-symmetry-dictated internal coordinates relaxed within a ferromagnetic configuration. While not the actual magnetic ground state, much recent experience~\cite{Sala2021,Pokharel2020PRL} finds this to be a much better approximation to actual structures than, for example, a non-spin polarized calculation, which inevitably neglects potential magnetoelastic effects~\cite{Pokharel2020PRL,Pokharel2018PRB,Yan2020PRM}. For all these calculations, an RK$_{max}$ value of $8.0$ was employed, where RK$_{max}$ is the product of the smallest muffin tin radius - in this case, Co, at $2.14$\,Bohr - and the maximum plane-wave expansion wavevector. The radius for the Y sphere was $2.5$\,Bohr, and spin-orbit coupling was not included for these calculations.

\section{Results and Discussion}

\subsection{Crystal Structure}

\begin{center}
	\begin{table*}[!htb]
		\caption{Atomic Positions, Equivalent Displacement Parameters (Ueq), and Occupancy (Occ) for Y$_{2}$Co$_{3}$\label{1}}	
		\begin{ruledtabular}
			\begin{tabular}{c c c c c c c }
				atom & site & x & y & z & U$_{eq}$ & Occ \\ \hline
				Y  & 8f & 0 & 0.16304 & 0.09760 & 0.015 & 1 \\ 
				Co1 & 8e & 1/4 & 0.41353 & 1/4 & 0.016 & 1 \\ 
				Co2 & 4b & -1/4 & 0    & 0  & 0.015 & 1 \\ 	
			\end{tabular}
		\end{ruledtabular}
	\end{table*}
\end{center}

Single crystal XRD analysis shows that Y$_{2}$Co$_{3}$ crystallizes in an orthorhombic structure instead of previously reported cubic structure~\cite{pelleg1965yttrium}. The space group is \textit{Cmce} (No. 64), La$_{2}$Ni$_{3}$-type structure, with cell parameters \textit{a} = 5.3302(11)\,\AA, \textit{b} = 9.5067(19)\,\AA\ and \textit{c} = 7.1127(14)\,\AA. Table~\ref{1} gives the atomic position (x, y, z), equivalent parameters and occupancy for Y$_{2}$Co$_{3}$. Table~\ref{3} shows the crystallographic data and XRD refinement parameters. The crystal structure is isomorphic to that of La$_{2}$Co$_{3}$, \textit{a} = 4.853\,\AA, \textit{b} = 10.350\,\AA~ and \textit{c} = 7.801\,\AA ~\cite{Gignoux1985PB}).

\begin{center}
	\begin{table}[!htb]
		\caption{Crystallographic data and refinement parameters for Y$_{2}$Co$_{3}$. Deposition Number on CSD: 2099730.\label{3}}	
		\begin{ruledtabular}
			\scriptsize
			\begin{tabular}{p{4cm} p{4cm}}
				Empirical formula & Y$_{2}$Co$_{3}$ \\
				Formula weight & 354.61 \\
				Temperature & 290(2) K \\
				Wavelength & 0.71073 Å \\
				Crystal system & Orthorhombic \\
				Space group & \textit{Cmce} \\
				Unit cell dimensions & a = 5.3302(11) Å  $\alpha$ = 90° \\
				& b = 9.5067(19) Å $\beta$= 90° \\
				&c = 7.1127(14) Å $\gamma$ = 90° \\
				Volume & 360.42(13) Å$^{3}$ \\
				Density (calculated) & 6.535 Mg/m$^{3}$ \\
				Absorption coefficient &  45.092 mm$^{-1}$ \\
				F(000) & 636 \\
				Crystal size & 0.147 x 0.131 x 0.088 mm$^{3}$ \\
				Theta range  & 4.287 to 27.524°. \\
				Index ranges & -6 $\leqslant$ h $\leqslant$ 6, -12 $\leqslant$ k $\leqslant$ 12, -9 $\leqslant$ l $\leqslant $9 \\
				Reflections collected & 2150 \\
				Independent reflections & 235 [R(int) = 0.0371] \\
				Completeness to theta = 25.242° & 100.0 \% \\
				Absorption correction & Semi-empirical from equivalents \\
				Max. and min. transmission & 0.0439 and 0.0133 \\
				Refinement method & Full-matrix least-squares on F$^{2}$\\
				Data / restraints / parameters & 235 / 0 / 16 \\
				Goodness-of-fit on F$^{2}$ & 1.270\\
				Final R indices [I>\modif{2$\sigma$(I)}] & R1 = 0.0192, wR2 = 0.0458 \\
				\modif{Final R indices [all data]} & R1 = 0.0194, wR2 = 0.0458 \\
				Extinction coefficient & n/a \\
				Largest diff. peak and hole & 0.552 and -0.827 e.Å$^{-3}$\\	
			\end{tabular}
		\end{ruledtabular}
	\end{table}
\end{center}

\begin{figure}[ht]
	\centering
	\includegraphics[width=\linewidth]{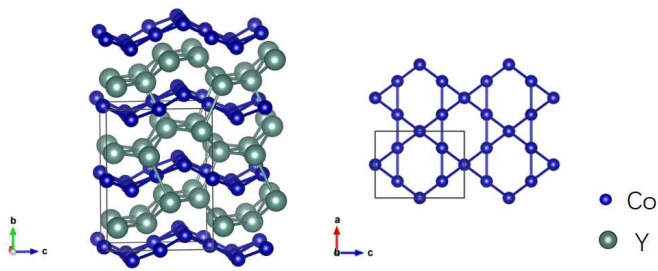} 
	\caption{View of crystal structure of Y$_{2}$Co$_{3}$.\label{structure}}  
\end{figure}

Figure~\ref{structure} shows an alternating layered Y and Co structure along the \textit{b} axis. In the \textit{ac} plane, the Co atoms form \modif{distorted Kagome structures with corner sharing triangles and hexagonal rings}. In La$_{2}$Co$_{3}$, the Co atoms in each layer form relatively regular \modif{Kagome nets}, with 2.427 or 2.471\,\AA\ for the distances between Co atoms and 119.4\degree\ or 121.2\degree\ for Co-Co-Co angles~\cite{Gignoux1985PB}. In Y$_{2}$Co$_{3}$, however, the Co-Co distances are 2.3692(4)\,\AA~ or 2.6651(5)\,\AA\ and the angles for Co-Co-Co are 139.40(3)\degree\ and 124.225(8)\degree. As a result, the \modif{Kagome lattices} in Y$_2$Co$_3$ are more distorted than in La$_2$Co$_3$. Table~\ref{2} lists the cell parameters of the R$_{2}$Co$_{3-x}$\{Si, Ga\}$_{x}$ system (R=La, Pr, Nd, Sm, Gd). With the increase of atomic number, the unit cell shrinks. The ratio \textit{a}/\textit{b} increases and \textit{c}/\textit{b} remains approximately the same, which suggests that the hexagonal rings in the \textit{ac} plane are stretched along the \textit{a} axis. The size of Y$^{3+}$ is smaller than that of Gd$^{3+}$, thus the distortion is even larger.

\begin{figure}[ht]
	\centering
	\includegraphics[width=\linewidth]{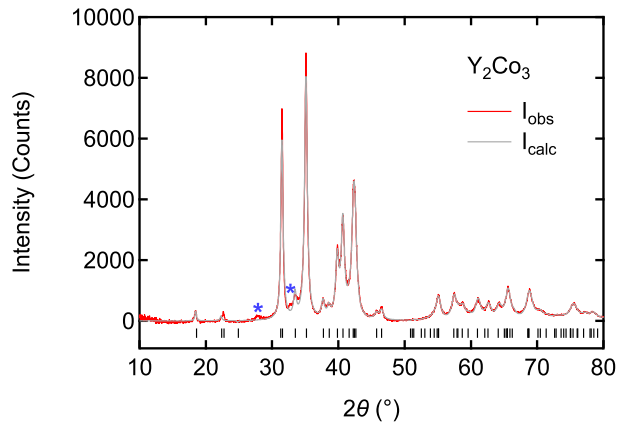} 
	\caption{\modif{X-Ray powder patterns (red line) and fits (grey line) of single crystal Y$_{2}$Co$_{3}$. The black bars indicate the peak locations expected from the single crystal XRD refinement at 290\,K. Peaks marked by * are from the impurity phase Y$_3$Co$_2$}.\label{PXRD}}  
\end{figure}

\modif{Figure~\ref{PXRD} shows the PXRD patterns with the Rietveld refinement. Peaks marked by ``$*$" are from the Y$_3$Co$_2$ impurity phase~\cite{Moreau1975ACB}, corresponding to the solidified flux on the surface of the single crystal samples.}

\begin{center}
	\begin{table*}[!ht]
		\caption{Unit cell parameters of the R$_{2}$Co$_{3-x}$\{Si, Ga\}$_{x}$ system.\label{2}}	
		\begin{ruledtabular}
			\begin{tabular}{c c c c c c c c }
				Compound & a (\AA) & b (\AA) & c (\AA) & V (\AA$^{3}$)& a/b & c/b & Ref \\ \hline
				La$_{2}$Co$_{3}$  & 4.853 & 10.350 & 7.801 & 391.83 & 0.4689 & 0.7537 &\cite{Gignoux1985PB} \\ 
				Pr$_{2}$Co$_{2.85}$Si$_{0.15}$ & 4.9064(3) & 10.0826(5)	& 7.6451(5)	& 378.2	& 0.4866 & 0.7582 & \cite{Mahon2018JAC} \\
				Nd$_{2}$Co$_{3}$ & 5.007(2)	& 9.981(3) & 7.519(2) & 375.76 & 0.5016 & 0.7533 & \cite{ray1973revised} \\
				Sm$_{2.1}$Co$_{2.65}$Si$_{0.25}$ & 5.3045(7) & 9.6625(1) & 7.2229(1) & 370.21 &	0.549 & 0.7475 & \cite{Mahon2018JAC} \\
				Gd$_{2}$Co$_{2.94}$Ga$_{0.06}$ & 5.315(3) & 9.613(4) & 7.169(5) & 366.29 & 0.5529 & 0.7458 & \cite{Sichevich1984Dopovidi} \\	
				Y$_{2}$Co$_{3}$	& 5.3302(11) & 9.5067(19) & 7.1127(14) & 360.42 & 0.5607 & 0.7481 & this work \\
			\end{tabular}
		\end{ruledtabular}
	\end{table*}
\end{center}

\subsection{Magnetic Properties}

\begin{figure}[!htb]
	\centering
	\includegraphics[width=\linewidth]{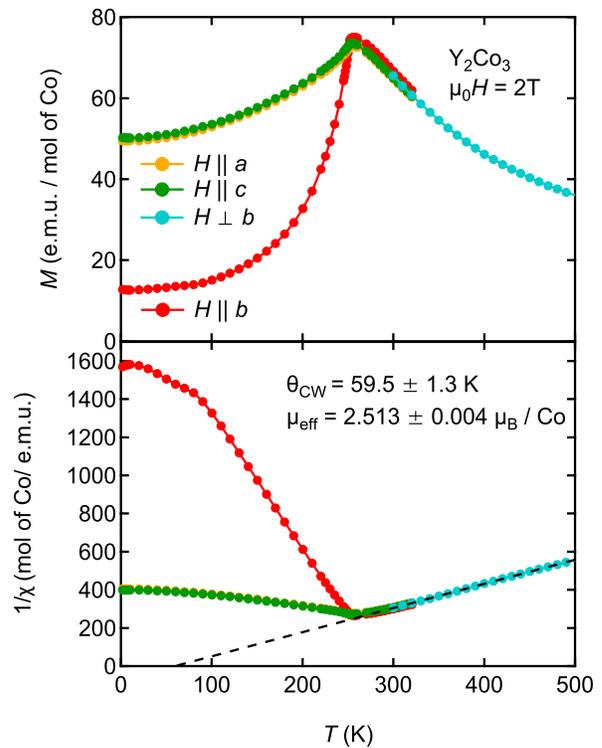} 
	\caption{(a) Magnetization and (b) inverse magnetic susceptibility of Y$_{2}$Co$_{3}$ as a function of temperature in an applied field of $2$\,T along the \textit{a}, \textit{b} and \textit{c} axes. \modif{The high temperature magnetization from 300 K to 500 K was measured using oven with magnetic field along the $ac$ plane.}\label{MvT}}  
\end{figure}

Figure~\ref{MvT} shows the temperature dependence of magnetization \textit{M} and inverse susceptibility with magnetic field of $2$\,T along the \textit{a}, \textit{b} and \textit{c} directions. A transition from antiferromagnetism to paramagnetism happens at \textit{T}$_{N}$ = 252\,K. A strong anisotropic behavior is observed. When the magnetic field is along the \textit{b} axis, the magnetization shows the sharpest transition, indicating that the magnetic moments primarily align along the \textit{b} axis. \modif{The effective moment $\mu_{eff}=2.513(4)$\,$\mu_{B}$/Co in the $ab$ plane and Curie-Weiss temperature $\theta_{CW}=59.5\pm1.3$\,K are obtained by fitting the curve above $T_{N}$ with the Curie-Weiss law. The Curie-Weiss temperature is relatively small, and the positive value confirms that there is weak ferromagnetic interaction.}

\begin{figure}[!htb]
	\centering
	\includegraphics[width=\linewidth]{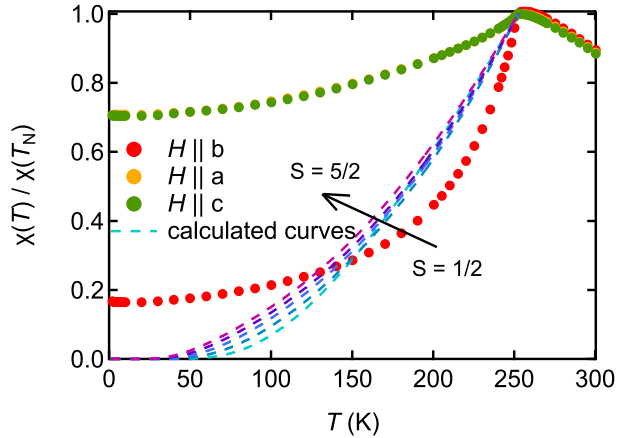} 
	\caption{Normalized susceptibility as a function of temperature. Dashed lines are the calculated curves in eqn.~\ref{eqn:Johnston}. with different values of $S$.\label{fitting}}  
\end{figure}

We compare the data below the N\'{e}el temperature following D.~Johnston's work~\cite{Johnston2012PRL} with the following equation:
\begin{equation} 
\frac{\chi_{||}(T)}{\chi(T_{N})}=\frac{1-f}{\tau^{\ast}-f},  \tau^{\ast}(t)=\frac{(S+1)t}{3SB'_{S}(y_{0})} \label{eqn:Johnston}
\end{equation}
where $\chi_{||}$ is the colinear parallel susceptibility, $t=\frac{T}{T_{N}}$, $f=\frac{\theta_{CW}}{T_{N}}=-0.08$ is calculated from the experiment data,  $B'_{S}(y_{0})=dB_{S}(y)/dy|_{y=y_{0}}$ is the derivative of the Brillouin function, $y_{0}=\frac{3S\bar{\mu} _{0}}{(S+1)t}$, \modif{S is the spin state of Co ions, }and $\bar{\mu}_{0}$ is calculated from $\bar{\mu}_{0}=B_{S}(y_{0})$.
Figure \ref{fitting} shows the calculated collinear susceptibility along the axial direction compared with the experimental data. The calculated result does not fit the experimental data well, which suggests that the magnetic order in Y$_{2}$Co$_{3}$ is not a simple collinear antiferromagnetic order along the $b$ axis. For comparison, La$_{2}$Co$_{3}$ also has a non-collinear antiferromagnetic order. In La$_{2}$Co$_{3}$, the magnetic moment of the Co atoms on the two sites are different: on 4a sites, the spins are along the $c$ axis, with $\left|M_{c}\right| = 0.35 \pm 0.05\,\mu_{B}$. On 8e sites, the spins tilt away from the $c$ axis towards the $a$ axis, with $\left|M_{c}\right| = 0.85 \pm 0.05\,\mu_{B}$ and $\left|M_{a}\right| = 0.34 \pm 0.05\,\mu_{B}$\cite{Gignoux1985PB}. Such planar spin alignment along the \textit{ac} plane, however, is obviously different from that of Y$_{2}$Co$_{3}$. In Y$_{2}$Co$_{3}$, the spin alignment seems to be more complicated. Based on the magnetization anisotropy, Y$_{2}$Co$_{3}$ is likely to have a combination of axial alignments along the \textit{b} axis with some planar alignments in the \textit{ac} plane. Neutron scattering experiments are necessary to fully determine the magnetic structure.

\begin{figure}[!htb]
	\centering
	\includegraphics[width=\linewidth]{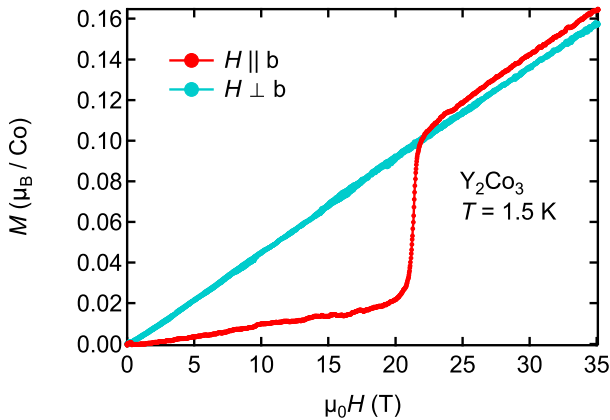} 
	\caption{\modif{Magnetization as a function of magnetic field at $T = 1.5$\,K. A spin flip transition is observed at 21 T when the magnetic field is along the $b$ direction.}\label{MvH}}  
\end{figure}

Figure~\ref{MvH} shows the magnetization of Y$_{2}$Co$_{3}$ as a function of magnetic field at \modif{$1.5$\,K. The magnetization increases linearly up to 21\,T, and a spin-flop transition is observed at $\mu_0H_{SF} = 21$\,T, indicating that the anisotropy energy is relatively weak compared to the antiferromagnetic exchange coupling energy. The magnetization does not saturate with applied magnetic field up to 35 T. This indicates that the antiferromagnetic ordering is robust and hard to suppress with the application of magnetic field. With the spin-flop transition field $H_{sf}$ and the slope of the $M$ vs $H$ curve above $H_{sf}$, one can estimate the magnetocrystalline anisotropy energy. The total energy per pair of Co of the spin-flop state can be written as~\cite{Blundell2001Magnetism}:
\begin{equation}
E_{sf}=-2M_sB\cos{\theta}+AM_s^2\cos{2\theta}-K\cos^2{\theta}\label{eqn:Blundell}
\end{equation}
where $B$ is the applied magnetic field along the direction where the magnetic moments spontaneous align with, $A$ is the constant connected with the exchange coupling, $K$ is the magnetocrystalline energy per pair of Co and $\theta$ is the angle between the magnetic moment and the direction of the magnetic field. The energy is minimized when $\cos{\theta}=\frac{M_sB}{2AM_s^2-K}$. The saturation magnetic field $B_{sat}$ can be obtained when $\theta=0\degree$, which yields $M_sB_{sat}=2AM_s^2-K$. Substituting this back to Eq.~\ref{eqn:Blundell}, one obtains the energy of spin-flop state as a function of magnetic field:
\begin{equation}
E_{sf}=-\frac{(M_sB)^2}{B_{sat}M_s}-AM_s^2\label{ESF}
\end{equation}
The spin-flop transition happens when the spin-flop energy equals the energy of the antiferromagnetic state $-AM_s^2-K$~\cite{Blundell2001Magnetism}. 
\begin{equation}
    -\frac{(M_sB_{sf})^2}{B_{sat}M_s}-AM_s^2=-AM_s^2-K
\end{equation}
This yields 
\begin{equation}
    K=\frac{(M_sB_{sf})^2}{B_{sat}M_s}=\frac{B_{sf}^2}{B_{sat}/M_s}\label{eqn:K}
\end{equation}
In Y$_2$Co$_3$, the saturation magnetic moment and magnetic field are not available, however, $B_{sat}/M_s$ in Eq.~\ref{eqn:K} can be calculated with the slope of the $M$ vs $H$ curve along the $b$ axis above the spin-flop transition field. Thus the magnetic anisotropy of Y$_2$Co$_3$ can be estimated to be $5.73$\,J/mol-Co or $4.85\times10^{-2}$\,J/g.} For comparison, Co-based antiferromagnets such as CaCo$_{2}$As$_{2}$ has two successive spin-flop transitions with much lower applied magnetic field at 3.5\,T and 4.7\,T due to the competition between exchange energy, magnetocrystalline anisotropy energy, and Zeeman energy and the magnetic anisotropy was reported as $3.76\times10^{-2}$\, J/g~\cite{Cheng2012PRB}, of the same scale as that of Y$_2$Co$_3$. Co$_{10}$Ge$_{3}$O$_{16}$ with $T_\textrm{N}=203$\,K shows complicated metamagnetic behavior depending on both temperature and magnetic field, and the metamagnetic transition was first observed at a temperature of 180 K and magnetic field of 3.9\,T~\cite{Barton2013PRB}, hydrogen containing compound such as Y$_{2}$Co$_{7}$H$_{6}$ shows spin-flip transitions with magnetic field of around 2\,T~\cite{bartashevich1983high}. On the other hand, the metamagnetic field can be very high, for example $80$\,T in K$_2$CoF$_4$~\cite{goto1992field}, $29$\,T in YCo$_3$H$_4$~\cite{bartashevich1994ultrahigh} and $14$\,T in YCo$_{3}$H$_{3.4}$~\cite{bartashevich1994ultrahigh}.

\begin{figure}[!htb]
	\centering
	\includegraphics[width=\linewidth]{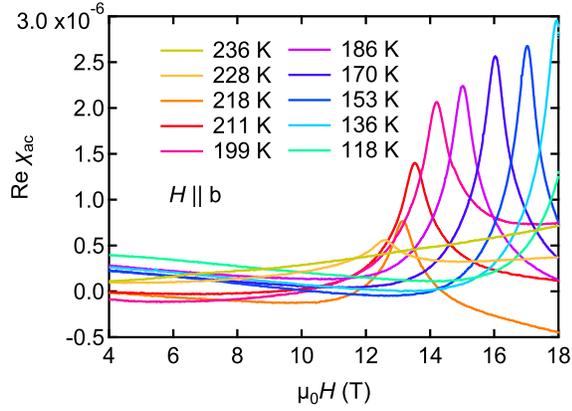} 
	\caption{\modif{AC susceptibility as a function of magnetic field at various temperatures.}\label{ACSusceptibility}}  
\end{figure}

\begin{figure}[!ht]
	\centering
	\includegraphics[width=\linewidth]{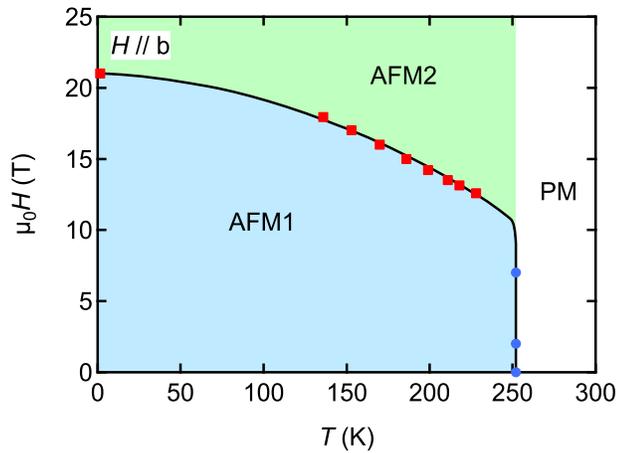} 
	\caption{\modiff{Magnetic phase diagram of Y$_2$Co$_3$ with the field along the $b$ axis. The boundary between the AFM2 state and the PM state is speculated. Data from field sweep and temperature sweep are shown as red squares and blue circles respectively.}\label{PhaseDiagram}}  
\end{figure}

\modif{Figure~\ref{ACSusceptibility}  }exhibits the AC susceptibility of Y$_2$Co$_3$ as a function of magnetic field along the $b$ direction at various temperatures from 1.5 K to 235 K. The large peaks are corresponding to the spin-flop transition. Figure~\ref{PhaseDiagram} shows the magnetic phase diagram of Y$_2$Co$_3$. The AFM1 state is the zero-field magnetic state where the magnetic moments primarily aligns along the $b$ direction and the AFM2 state is corresponding to a new magnetic structure due to the field-induced spin-flop transition.
\subsection{High Pressure Behavior}

\begin{figure}[!ht]
	\centering
	\includegraphics[width=\linewidth]{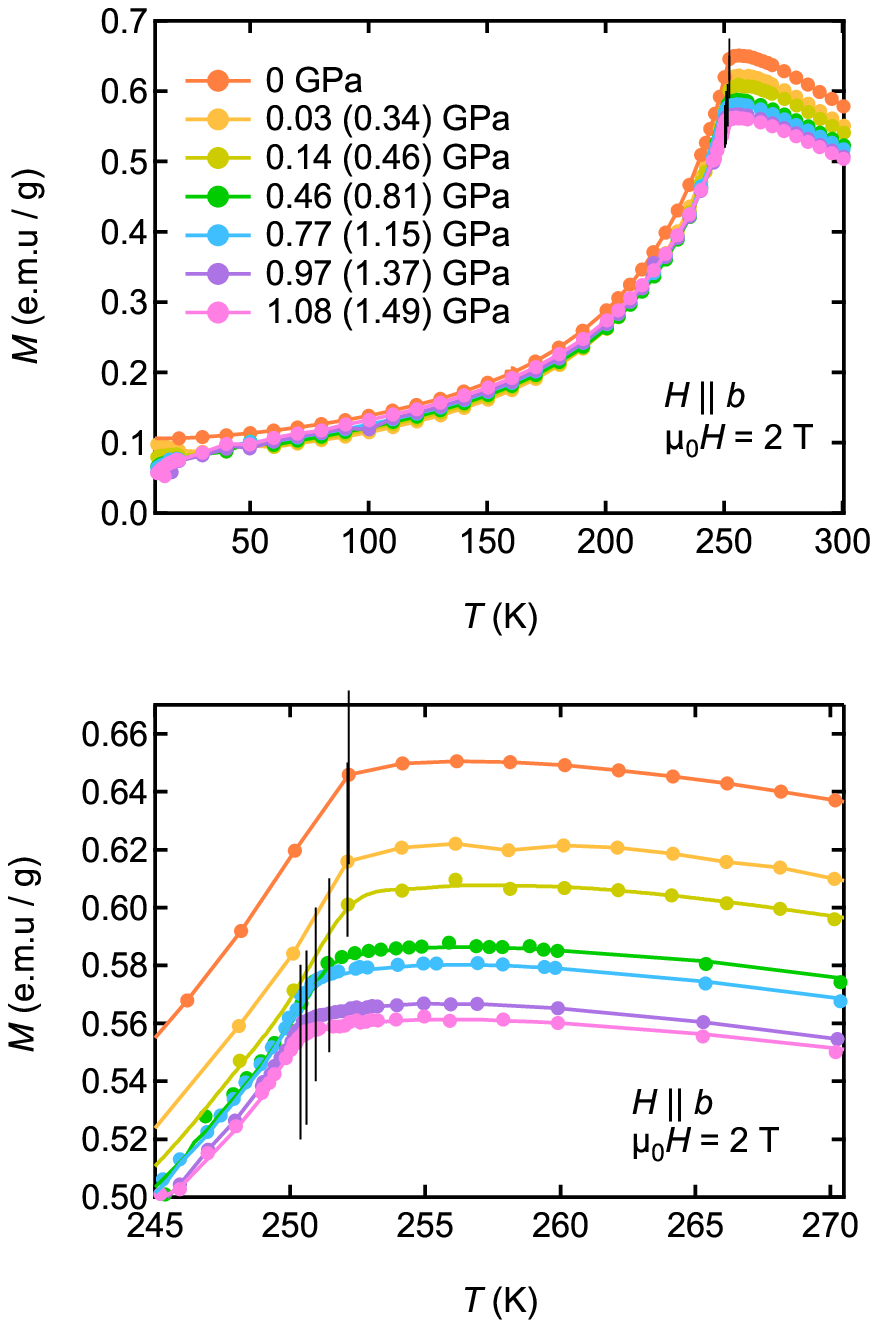} 
	\caption{The temperature dependence of the magnetization of Y$_{2}$Co$_{3}$ under various pressures. The pressures are determined at low temperature using the superconducting transition of Pb (and the values near room temperature are estimated with the ferromagnetic transition of Gd).}\label{P}  
\end{figure}

\begin{figure}[!ht]
	\centering
	\includegraphics[width=\linewidth]{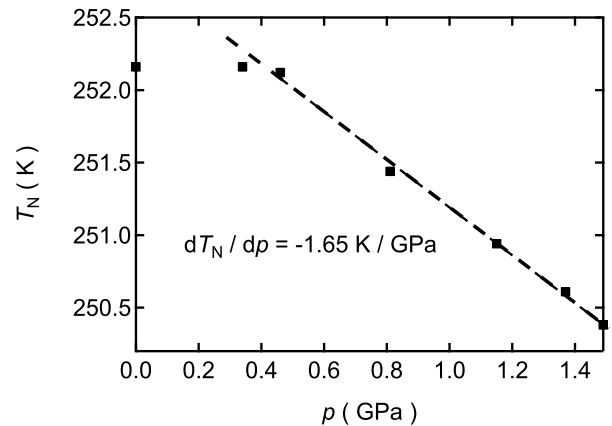} 
	\caption{The pressure dependence of the N\'{e}el temperature \textit{T}$_{N}$ for Y$_{2}$Co$_{3}$.
	}\label{TP}  
\end{figure}

To investigate the stability of the antiferromagnetic ordering in Y$_{2}$Co$_{3}$, we performed magnetization measurements of Y$_{2}$Co$_{3}$ up to 1\,GPa. As the pressure increases, $T_\textrm{N}$ of Y$_{2}$Co$_{3}$ decreases, as is shown in Figs.~\ref{P} and~\ref{TP}, with $dT_\textrm{N}/dp=-1.65$\,K/GPa. The antiferromagnetic ordering is not suppressed under pressure up to $1$\,GPa. Higher pressure with diamond anvil cells are necessary to further study the robustness of the antiferromagnetic order. 

\subsection{Electrical Resistivity}

\begin{figure}[!ht]
	\centering
	\includegraphics[width=\linewidth]{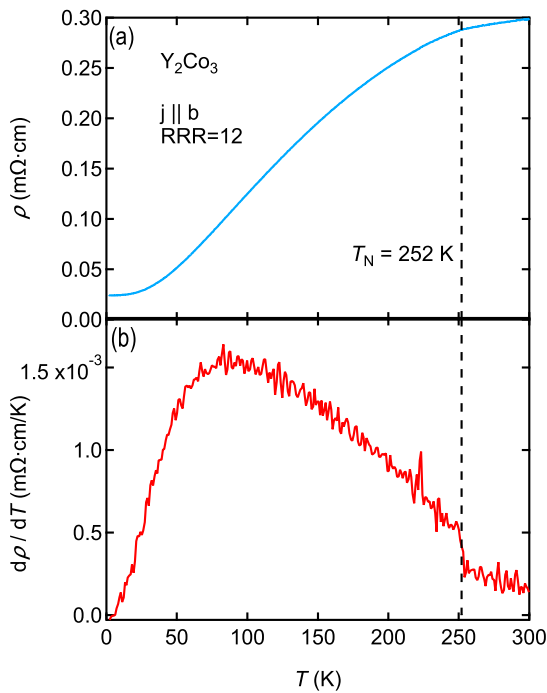} 
	\caption{(a) Electrical resistivity and (b) temperature derivative of the resistivity as a function of temperature for Y$_{2}$Co$_{3}$.\label{R}}  
\end{figure}

The temperature dependence of electrical resistivity along the \textit{b} axis of Y$_{2}$Co$_{3}$ is shown in figure~\ref{R}(a). A metallic behavior is observed below 300\,K. At 252\,K, there is a discontinuous change of slope of the resistivity, which is determined by a step in $d\rho/dT$ (figure~\ref{R}(b)). This corresponds to the antiferromagnetic transition.

\subsection{Heat Capacity}

\begin{figure}[!ht]
	\centering
	\includegraphics[width=\linewidth]{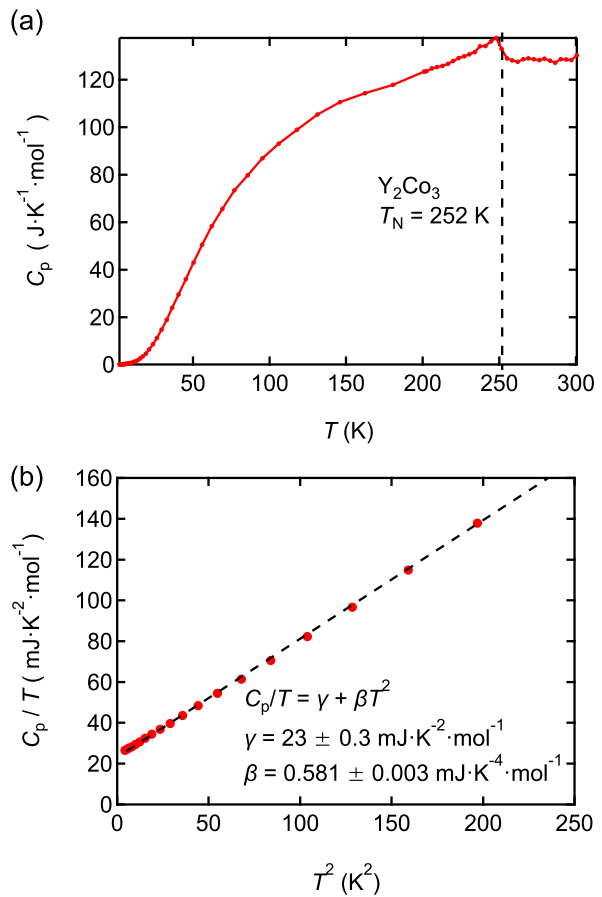} 
	\caption{(a)Temperature dependence of specific heat capacity $C_{p}$ for single crystal Y$_{2}$Co$_{3}$. (b) $C_{p}/T$ as a function of $T^{2}$. The black dash line is the fitting curve using the formula $C_{p}/T=\gamma+\beta T^{2}$. 
	}\label{HC}  
\end{figure}

Figure~\ref{HC}(a) shows the temperature dependence of the heat capacity of single crystalline Y$_{2}$Co$_{3}$. The peak at 252\,K indicates the antiferromagnetic phase transition, which is consistent with $T_\textrm{N}=252$\,K obtained from the magnetization and resistivity measurements. The finite jump of the heat capacity at the N\'{e}el temperature is about 10\,J\,K$^{-1}$\,mol$^{-1}$, indicating a second order phase transition in relatively good agreement with the classical mean field theory where $\Delta C_p$ at the transition is given by $\frac{3}{2}NR=37.4$\,J\,K$^{-1}$\,mol$^{-1}$ with $N=3$ the number of Co in the formula unit~\cite{wipf2012statistical}. Figure~\ref{HC}(b) shows the linear relationship between $C_{p}/T$ and $T^{2}$. Following the relation $C_{p}/T=\gamma+\beta T^{2}$, we obtain $\gamma=23.0(3)$\,mJ\,mol$^{-1}$\,K$^{-2}$ and $\beta=0.581(3)$\,mJ\,mol$^{-1}$\,K$^{-4}$. The Debye temperature $T_\textrm{D}=254$\,K is obtained from $T_\textrm{D}=(\frac{12\pi^{4}NR}{5\beta})^\frac{1}{3}$~\cite{kittel1996introduction}, where $N=5$ is the number of atoms in the chemical formula and $R=8.314$\,J\,mol$^{-1}$\,K$^{-1}$ is the gas constant. 

\subsection{First Principles Calculations}

Any attempt to develop a full picture of the magnetic interactions in this material should begin with a careful look at its complex physical structure, as in Fig.~\ref{structure}. While there are just two crystallographically distinct Co sites, the zigzag nature of the Co layers, along with the complex hexagonal and triangular coordination of these layers, as in the right panel of Fig.~\ref{structure}, suggests that practitioners of first principles calculations are facing a formidable physical system to describe. Despite this, we have been able to make progress in its description.

The first experimental fact that first principles calculations should explain is its experimentally apparent antiferromagnetism. As Figure~\ref{Co compound} suggests, compounds with 60 or more atomic percent of Cobalt are much more often ferromagnetic than antiferromagnetic. While Nature delights in exceptions to such simple classifications, such as the surprisingly paramagnetic CeCo$_3$ \cite{Lamichhane2018PRA,Lamichhane2020PM,Pandey2018PRA}, one would naively expect antiferromagnetic behavior in this stoichiometry regime to be the province of electronegative anions such as Oxygen, Sulfur or the Fluorine group. Yttrium, by contrast, is known substantially for its mineral occurrence with the ``rare earth" family, and similar chemical properties, despite its lack of $f$ electrons. Y$_2$Co$_3$ is hardly a likely candidate for complex antiferromagnetic behavior, though the existence of such behavior in La$_2$Co$_3$ at least makes this finding more plausible.    

\begin{figure*}[ht]
	\centering
	\includegraphics[width=\linewidth]{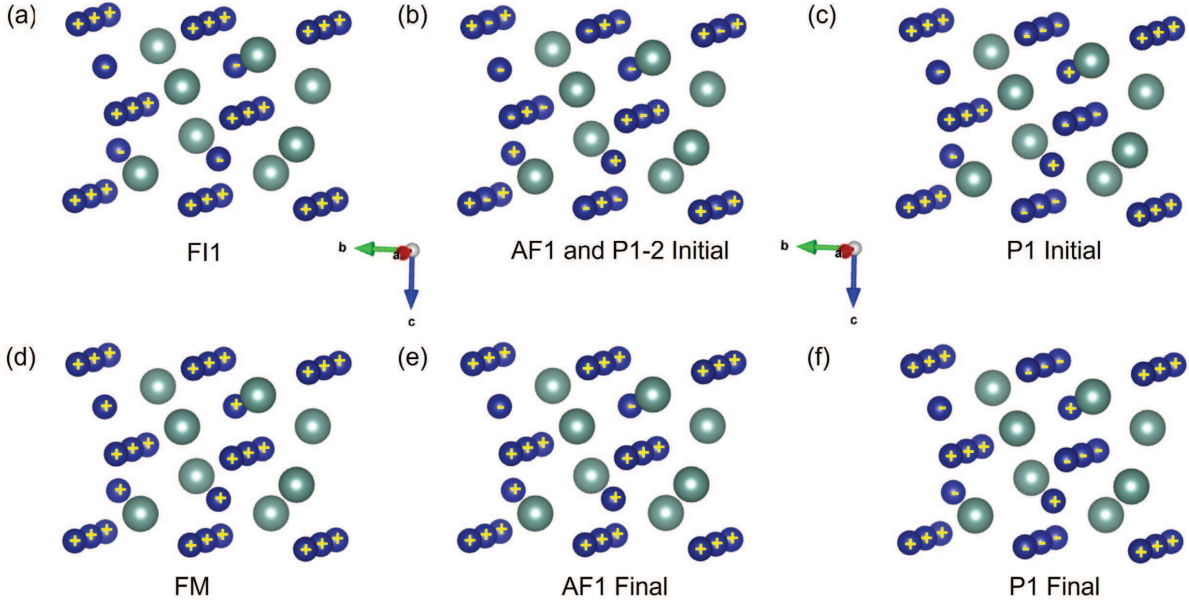} 
	\caption{\modif{The magnetic structure of FI1, AF1, P1-2 and P1 states. (a) The initial state of the FI1 model, which converges to a ferromagnetic state (d). (b) The initial state of AF1 and P1-2 models. AF1 state converge to a complex magnetic state (e) which is 54 meV/Co above the ferromagnetic state. The P1-2 model was performed in a lower symmetry, and the calculation failed to converge. (c) The initial state of the P1 model, which converges to an antiferromagnetic structure (f) and is energetically degenerate with the ferromagnetic configuration.}\label{calc_mag}}
\end{figure*}

Our calculations in fact find evidence, when combined with experimental information such as the substantial ordering temperature, for complex antiferromagnetic behavior. In addition to the ferromagnetic configuration, we also tried a {\it ferri}magnetic calculation with the two distinct Co sites anti-aligned (referred to as ``FI1"), as well as still more complex configurations in which each of the two distinct sites was broken into two separate sites and a fully antiferromagnetic configuration was initialized (``AF1"), shown in Fig.~\ref{calc_mag}. None of these calculations, however, produce a distinct magnetic state energetically competitive with the ferromagnetic solution (for which the fourfold and twofold Co sites had respective moments of $0.87$ and $1.42$\,$\mu_{B}$). 
The state initialized as ``FI1" ultimately converges to the ferromagnetic solution, suggestive of itinerant magnetic character. Supporting this assertion of itinerancy is the ultimate convergence of the ``AF1"-initialized calculation, not to an antiferromagnetic state, but to a complex magnetic state with respective moments on the split four-fold and two-fold sites of $0.16$, $0.16$, $-1.04$ and $1.28$\,$\mu_B$. This state, however, falls some $54$\,meV/Co above the ferromagnetic configuration, though this energy difference is at least 
roughly consistent with the ordering point of $252$\,K. The variability of the moment size relative to the ferromagnetic solution is also suggestive of itinerant character.

\begin{figure}[ht]
	\centering
	\includegraphics[width=0.85\linewidth]{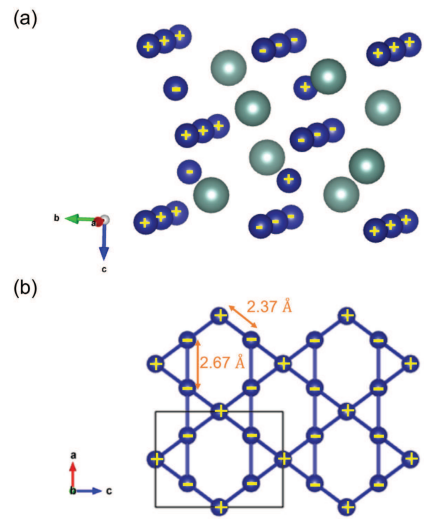} 
	\caption{(a) The magnetic structure of the P1 antiferromagnetic state. (b) The magnetic structure in the $ac$ plane.\label{Mag}} 
\end{figure}

Faced with this complicated situation, and the experimental finding of a complex antiferromagnetic state in La$_2$Co$_3$, we chose to study antiferromagnetic states in the full 4 formula unit cell with all 12 Co atoms considered as independent. This cell is of P1 symmetry and calculations are correspondingly protracted. Two specific states, among the manifold of possible states were studied, which we term ``P1" and ``P1-2". The initialization pattern of P1 is shown in Fig.~\ref{Mag}.

A rationale for the magnetic structure ``P1" can be understood as follows. In La$_2$Co$_3$, the shortest Co-Co distance is 2.43\,\AA, with a ferromagnetic interaction between these two atoms. In Y$_2$Co$_3$, however, the Co-Co separations are about 2.37\,\AA\, and 2.67\,\AA~ in the distorted hexagonal rings in the $ac$ plane, as is shown in figure~\ref{Mag}(b). Assuming there is a critical separation of $2.40(3)$\,\AA~ below which the direct interaction is antiferromagnetic, we obtain the magnetic structure in the $ac$~plane shown in figure~\ref{Mag}(b). Along the $b$ direction, the Co layers are separated by Y layers, and the antiferromagnetic alignments show up due to the super exchange.

This initialization pattern is in fact retained throughout the calculation, the moment magnitudes are identical (sign excepted) to that of the ferromagnetic solution, and the energy is degenerate, within calculational precision to the ferromagnetic solution.  The finding of identical moment magnitudes to the ferromagnetic solution is in fact suggestive of local, not itinerant character, and one may thus consider a complicated dual itinerant-local moment behavior, as has previously 
been observed for the parent compounds of the iron-based superconductors \cite{Moon2010PRB}. 

It is of interest to note that ``P1" is {\it not} a ``maximally" antiferromagnetic state, in the sense of having as many nearest and next-nearest neighbor Co-Co pairs anti-aligned. One may readily observe from Fig.~\ref{Mag} the existence of ferromagnetically coupled chains in the  
a direction, which is the nearest-neighbor Co-Co interaction. In the P1-2 state, we have set these nearest-neighbors to be antiferromagnetically coupled. However, this relative orientation switches in the course of the calculation, complex magnetic orientations ensue, and the calculation failed to converge after some 250 iterations. This level of calculational difficulty is characteristic of rare-earth compounds, yet Yttrium has no 4f electrons. Note also that the initialized ``P1" state ultimately converges to a state of orthorhombic (Pnma) symmetry but this was found after the fact so that for clarity we retain the original designation.

\begin{figure}[ht]
	\centering
	\includegraphics[width=0.75\linewidth,angle=270]{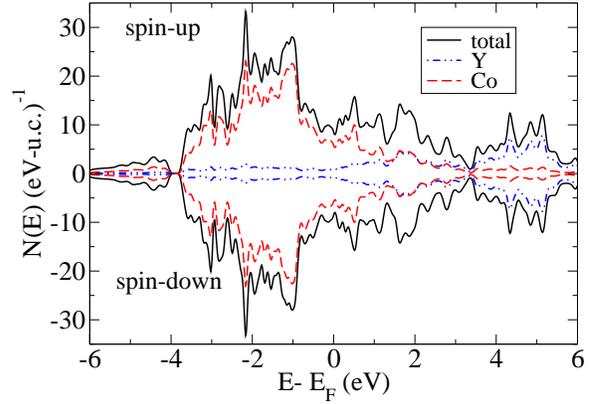} 
	\caption{The calculated density-of-states of the P1 antiferromagnetic state.\label{DOS}} 
\end{figure}

Note that while the ``P1" state is energetically degenerate with the ferromagnetic solution, there are numerous degrees of freedom allowing a substantial manifold of non-collinear magnetic states (not studied here) by which the system can very likely significantly lower its energy, in view of the obvious competition of ferromagnetic and antiferromagnetic interactions and the additional frustration associated with the triangular and hexagonal planar coordination. We therefore argue that the existence of the relatively low-lying antiferromagnetic P1 state is evidence for the complex magnetic character observed experimentally.
In Figure~\ref{DOS} we present the calculated density-of-states in the P1 state. As expected, most of the character is Cobalt, and the general symmetry of spin-up and spin-down DOS confirms the antiferromagnetic character of the P1 state.  From the Fermi-level density-of-states
we find a specific-heat $\gamma$ of $9$\,mJ/mol-K$^{2}$, much less than the experimental value. This could suggest strong electron-phonon coupling, but it remains an open theoretical question in view of the uncertainty in the actual magnetic structure). 

In summary, our calculations of the magnetism in Y$_2$Co$_3$ find evidence for a complex antiferromagnetic state, likely containing substantial components of both itinerant and local-moment character, deriving from a complicated, rather frustrated physical structure.

The antiferromagnetic inter-chain alignment (i.e. chains separated by half a b lattice spacing) is very likely associated with the presence of the intervening Y atom, while the antiferromagnetic alignment within the distorted hexagonal plane (see Figure~\ref{Mag}(b)), suggests that the effects of the Y atom are not limited to its immediate physical location, but pervade throughout the system, which is consistent with the finding of substantial itinerant character in this complex system.

\section {Conclusion}
In summary, we report on a solution growth method to synthesize single crystals of the new antiferromagnetic compound Y$_2$Co$_3$. Our study shows that Y$_{2}$Co$_{3}$ has a La$_{2}$Ni$_{3}$-type orthorhombic crystal  structure, with space group \textit{Cmce} (No.64). We find that Y$_{2}$Co$_{3}$ has a robust antiferromagnetic order below $T_\textrm{N}=252$\,K. Magnetization measurements show that the moments are aligned mostly along the $b$ axis, with a complex non-collinear magnetic structure, \modif{which is likely due to the distorted Kagome net of Co. High magnetic field measurements show a spin-flop behavior at 21 T. The antiferromagnetic order is not suppressed with the applied magnetic field up to 35 T}. The DFT calculations find evidence for a complex antiferromagnetic state, likely containing substantial components of both itinerant and local-moment character. \modif{However, the first principle calculations cannot easily capture the relatively high N\'{e}el temperature and stability at high-field.}

\section {Acknowledgment}

We thank Shanti Deemyad and Audrey Grockowiak for useful discussions. V.T. and Y.S. acknowledge support from the UC Lab Fees Research Program (LFR-20-653926) and UC Davis Startup funds. Work at Oak Ridge National Laboratory (first principles calculations) was supported by the U.S. Department of Energy, Office of Science, Basic Energy Sciences, Materials Science and Engineering Division. K.P.D. and S.M.K. acknowledge NSF DMR-1709382, -2001156 for funding. \modif{A portion of this work was performed at the National High Magnetic Field Laboratory, which is supported by the National Science Foundation Cooperative Agreement No. DMR-1644779 and the State of Florida.}

\bibliography{Y2Co3,biblio}

\begin{thebibliography}{39}
\expandafter\ifx\csname natexlab\endcsname\relax\def\natexlab#1{#1}\fi
\expandafter\ifx\csname bibnamefont\endcsname\relax
  \def\bibnamefont#1{#1}\fi
\expandafter\ifx\csname bibfnamefont\endcsname\relax
  \def\bibfnamefont#1{#1}\fi
\expandafter\ifx\csname citenamefont\endcsname\relax
  \def\citenamefont#1{#1}\fi
\expandafter\ifx\csname url\endcsname\relax
  \def\url#1{\texttt{#1}}\fi
\expandafter\ifx\csname urlprefix\endcsname\relax\def\urlprefix{URL }\fi
\providecommand{\bibinfo}[2]{#2}
\providecommand{\eprint}[2][]{\url{#2}}

\bibitem[{\citenamefont{H{\'{e}}bert et~al.}(2021)\citenamefont{H{\'{e}}bert,
  Barbier, Berthebaud, Lebedev, Pralong, and Maignan}}]{Hebert2021Transport}
\bibinfo{author}{\bibfnamefont{S.}~\bibnamefont{H{\'{e}}bert}},
  \bibinfo{author}{\bibfnamefont{T.}~\bibnamefont{Barbier}},
  \bibinfo{author}{\bibfnamefont{D.}~\bibnamefont{Berthebaud}},
  \bibinfo{author}{\bibfnamefont{O.~I.} \bibnamefont{Lebedev}},
  \bibinfo{author}{\bibfnamefont{V.}~\bibnamefont{Pralong}}, \bibnamefont{and}
  \bibinfo{author}{\bibfnamefont{A.}~\bibnamefont{Maignan}},
  \bibinfo{journal}{The Journal of Physical Chemistry C}
  \textbf{\bibinfo{volume}{125}}, \bibinfo{pages}{5386} (\bibinfo{year}{2021}),
  \urlprefix\url{https://doi.org/10.1021/acs.jpcc.0c11601}.

\bibitem[{\citenamefont{Menshikov et~al.}(1985)\citenamefont{Menshikov, Takzei,
  Dorofeev, Kazantsev, Kostyshin, and Sych}}]{menshikov1985magnetic}
\bibinfo{author}{\bibfnamefont{A.~Z.} \bibnamefont{Menshikov}},
  \bibinfo{author}{\bibfnamefont{G.~A.} \bibnamefont{Takzei}},
  \bibinfo{author}{\bibfnamefont{Y.~A.} \bibnamefont{Dorofeev}},
  \bibinfo{author}{\bibfnamefont{V.~A.} \bibnamefont{Kazantsev}},
  \bibinfo{author}{\bibfnamefont{A.~K.} \bibnamefont{Kostyshin}},
  \bibnamefont{and} \bibinfo{author}{\bibfnamefont{I.~I.} \bibnamefont{Sych}},
  \bibinfo{journal}{Zh. Eksp. Teor. Fiz.} \textbf{\bibinfo{volume}{89}},
  \bibinfo{pages}{1269} (\bibinfo{year}{1985}).

\bibitem[{\citenamefont{Hamm et~al.}(2016)\citenamefont{Hamm, G{\"{o}}lden,
  Hildebrandt, Weischenberg, Zhang, Alff, and Birkel}}]{Hamm2016}
\bibinfo{author}{\bibfnamefont{C.~M.} \bibnamefont{Hamm}},
  \bibinfo{author}{\bibfnamefont{D.}~\bibnamefont{G{\"{o}}lden}},
  \bibinfo{author}{\bibfnamefont{E.}~\bibnamefont{Hildebrandt}},
  \bibinfo{author}{\bibfnamefont{J.}~\bibnamefont{Weischenberg}},
  \bibinfo{author}{\bibfnamefont{H.}~\bibnamefont{Zhang}},
  \bibinfo{author}{\bibfnamefont{L.}~\bibnamefont{Alff}}, \bibnamefont{and}
  \bibinfo{author}{\bibfnamefont{C.~S.} \bibnamefont{Birkel}},
  \bibinfo{journal}{Journal of Materials Chemistry C}
  \textbf{\bibinfo{volume}{4}}, \bibinfo{pages}{7430} (\bibinfo{year}{2016}),
  ISSN \bibinfo{issn}{20507526}, \urlprefix\url{www.rsc.org/MaterialsC}.

\bibitem[{\citenamefont{Wu et~al.}(1991)\citenamefont{Wu, Chuang, and
  Su}}]{wu1991re}
\bibinfo{author}{\bibfnamefont{C.}~\bibnamefont{Wu}},
  \bibinfo{author}{\bibfnamefont{Y.}~\bibnamefont{Chuang}}, \bibnamefont{and}
  \bibinfo{author}{\bibfnamefont{X.}~\bibnamefont{Su}},
  \bibinfo{journal}{Zeitschrift f{\"u}r Metallkunde}
  \textbf{\bibinfo{volume}{82}}, \bibinfo{pages}{73} (\bibinfo{year}{1991}).

\bibitem[{\citenamefont{Pelleg and Carlson}(1965)}]{pelleg1965yttrium}
\bibinfo{author}{\bibfnamefont{J.}~\bibnamefont{Pelleg}} \bibnamefont{and}
  \bibinfo{author}{\bibfnamefont{O.}~\bibnamefont{Carlson}},
  \bibinfo{journal}{Journal of the Less Common Metals}
  \textbf{\bibinfo{volume}{9}}, \bibinfo{pages}{281} (\bibinfo{year}{1965}).

\bibitem[{\citenamefont{White}(1992)}]{white1992magnetic}
\bibinfo{author}{\bibfnamefont{M.}~\bibnamefont{White}}, \bibinfo{type}{Tech.
  Rep.}, \bibinfo{institution}{State Univ. of New York, Buffalo, NY (United
  States)} (\bibinfo{year}{1992}).

\bibitem[{\citenamefont{Byland et~al.}(2021)\citenamefont{Byland, Shi
  et~al.}}]{Byland}
\bibinfo{author}{\bibfnamefont{J.~K.} \bibnamefont{Byland}},
  \bibinfo{author}{\bibfnamefont{Y.}~\bibnamefont{Shi}}, \bibnamefont{et~al.},
  \bibinfo{journal}{In preparation}  (\bibinfo{year}{2021}).

\bibitem[{\citenamefont{Ray et~al.}(1973)\citenamefont{Ray, Biermann, Harmer,
  and Davison}}]{ray1973revised}
\bibinfo{author}{\bibfnamefont{A.}~\bibnamefont{Ray}},
  \bibinfo{author}{\bibfnamefont{A.}~\bibnamefont{Biermann}},
  \bibinfo{author}{\bibfnamefont{R.}~\bibnamefont{Harmer}}, \bibnamefont{and}
  \bibinfo{author}{\bibfnamefont{J.}~\bibnamefont{Davison}},
  \bibinfo{type}{Tech. Rep.} (\bibinfo{year}{1973}).

\bibitem[{\citenamefont{Gignoux et~al.}(1985)\citenamefont{Gignoux, Lemaire,
  Mendia-Monterroso, Moreau, and Schweizer}}]{Gignoux1985PB}
\bibinfo{author}{\bibfnamefont{D.}~\bibnamefont{Gignoux}},
  \bibinfo{author}{\bibfnamefont{R.}~\bibnamefont{Lemaire}},
  \bibinfo{author}{\bibfnamefont{R.}~\bibnamefont{Mendia-Monterroso}},
  \bibinfo{author}{\bibfnamefont{J.}~\bibnamefont{Moreau}}, \bibnamefont{and}
  \bibinfo{author}{\bibfnamefont{J.}~\bibnamefont{Schweizer}},
  \bibinfo{journal}{Physica B+C} \textbf{\bibinfo{volume}{130}},
  \bibinfo{pages}{376} (\bibinfo{year}{1985}), ISSN \bibinfo{issn}{0378-4363}.

\bibitem[{\citenamefont{Tenc{\'{e}} et~al.}(2014)\citenamefont{Tenc{\'{e}},
  {Caballero Flores}, Chable, Gorsse, Chevalier, and
  Gaudin}}]{Tence2014Stabilization}
\bibinfo{author}{\bibfnamefont{S.}~\bibnamefont{Tenc{\'{e}}}},
  \bibinfo{author}{\bibfnamefont{R.}~\bibnamefont{{Caballero Flores}}},
  \bibinfo{author}{\bibfnamefont{J.}~\bibnamefont{Chable}},
  \bibinfo{author}{\bibfnamefont{S.}~\bibnamefont{Gorsse}},
  \bibinfo{author}{\bibfnamefont{B.}~\bibnamefont{Chevalier}},
  \bibnamefont{and} \bibinfo{author}{\bibfnamefont{E.}~\bibnamefont{Gaudin}},
  \bibinfo{journal}{Inorganic Chemistry} \textbf{\bibinfo{volume}{53}},
  \bibinfo{pages}{6728} (\bibinfo{year}{2014}), ISSN \bibinfo{issn}{1520510X}.

\bibitem[{\citenamefont{Mahon et~al.}(2018)\citenamefont{Mahon, Gaudin,
  Vignolle, Ballon, Chevalier, and Tenc\'e}}]{Mahon2018JAC}
\bibinfo{author}{\bibfnamefont{T.}~\bibnamefont{Mahon}},
  \bibinfo{author}{\bibfnamefont{E.}~\bibnamefont{Gaudin}},
  \bibinfo{author}{\bibfnamefont{B.}~\bibnamefont{Vignolle}},
  \bibinfo{author}{\bibfnamefont{G.}~\bibnamefont{Ballon}},
  \bibinfo{author}{\bibfnamefont{B.}~\bibnamefont{Chevalier}},
  \bibnamefont{and} \bibinfo{author}{\bibfnamefont{S.}~\bibnamefont{Tenc\'e}},
  \bibinfo{journal}{Journal of Alloys and Compounds}
  \textbf{\bibinfo{volume}{737}}, \bibinfo{pages}{377} (\bibinfo{year}{2018}),
  ISSN \bibinfo{issn}{0925-8388},
  \urlprefix\url{http://www.sciencedirect.com/science/article/pii/S0925838817341889}.

\bibitem[{\citenamefont{Morozkin et~al.}(2015)\citenamefont{Morozkin, Isnard,
  Nirmala, and Malik}}]{Morozkin2015JSSC}
\bibinfo{author}{\bibfnamefont{A.~V.} \bibnamefont{Morozkin}},
  \bibinfo{author}{\bibfnamefont{O.}~\bibnamefont{Isnard}},
  \bibinfo{author}{\bibfnamefont{R.}~\bibnamefont{Nirmala}}, \bibnamefont{and}
  \bibinfo{author}{\bibfnamefont{S.~K.} \bibnamefont{Malik}},
  \bibinfo{journal}{Journal of Solid State Chemistry}
  \textbf{\bibinfo{volume}{225}}, \bibinfo{pages}{368} (\bibinfo{year}{2015}),
  ISSN \bibinfo{issn}{0022-4596},
  \urlprefix\url{http://www.sciencedirect.com/science/article/pii/S0022459615000250}.

\bibitem[{\citenamefont{Canfield and Fisher}(2001)}]{canfield2001high}
\bibinfo{author}{\bibfnamefont{P.~C.} \bibnamefont{Canfield}} \bibnamefont{and}
  \bibinfo{author}{\bibfnamefont{I.~R.} \bibnamefont{Fisher}},
  \bibinfo{journal}{Journal of Crystal Growth} \textbf{\bibinfo{volume}{225}},
  \bibinfo{pages}{155} (\bibinfo{year}{2001}).

\bibitem[{\citenamefont{Jesche and Canfield}(2014)}]{Jesche2014PM}
\bibinfo{author}{\bibfnamefont{A.}~\bibnamefont{Jesche}} \bibnamefont{and}
  \bibinfo{author}{\bibfnamefont{P.~C.} \bibnamefont{Canfield}},
  \bibinfo{journal}{Philosophical Magazine} \textbf{\bibinfo{volume}{94}},
  \bibinfo{pages}{2372} (\bibinfo{year}{2014}).

\bibitem[{\citenamefont{Sheldrick}(2008)}]{sheldrick2008short}
\bibinfo{author}{\bibfnamefont{G.~M.} \bibnamefont{Sheldrick}},
  \bibinfo{journal}{Acta Crystallographica Section A: Foundations of
  Crystallography} \textbf{\bibinfo{volume}{64}}, \bibinfo{pages}{112}
  (\bibinfo{year}{2008}).

\bibitem[{\citenamefont{Jesche et~al.}(2016)\citenamefont{Jesche, Fix,
  Kreyssig, Meier, and Canfield}}]{jesche2016x}
\bibinfo{author}{\bibfnamefont{A.}~\bibnamefont{Jesche}},
  \bibinfo{author}{\bibfnamefont{M.}~\bibnamefont{Fix}},
  \bibinfo{author}{\bibfnamefont{A.}~\bibnamefont{Kreyssig}},
  \bibinfo{author}{\bibfnamefont{W.~R.} \bibnamefont{Meier}}, \bibnamefont{and}
  \bibinfo{author}{\bibfnamefont{P.~C.} \bibnamefont{Canfield}},
  \bibinfo{journal}{Philosophical magazine} \textbf{\bibinfo{volume}{96}},
  \bibinfo{pages}{2115} (\bibinfo{year}{2016}).

\bibitem[{\citenamefont{Bartholin and Bloch}(1968)}]{Bartholin1968}
\bibinfo{author}{\bibfnamefont{H.}~\bibnamefont{Bartholin}} \bibnamefont{and}
  \bibinfo{author}{\bibfnamefont{D.}~\bibnamefont{Bloch}},
  \bibinfo{journal}{Journal of Physics and Chemistry of Solids}
  \textbf{\bibinfo{volume}{29}}, \bibinfo{pages}{1063} (\bibinfo{year}{1968}),
  ISSN \bibinfo{issn}{00223697}.

\bibitem[{\citenamefont{Bireckoven and Wittig}({1988})}]{Bireckoven1988JPESI}
\bibinfo{author}{\bibfnamefont{B.}~\bibnamefont{Bireckoven}} \bibnamefont{and}
  \bibinfo{author}{\bibfnamefont{J.}~\bibnamefont{Wittig}},
  \bibinfo{journal}{{J. Phys. E: Sci. Instrum.}}
  \textbf{\bibinfo{volume}{{21}}}, \bibinfo{pages}{841}
  (\bibinfo{year}{{1988}}), ISSN \bibinfo{issn}{{0022-3735}}.

\bibitem[{\citenamefont{Blaha et~al.}(2001)\citenamefont{Blaha, Schwarz,
  Madsen, Kvasnicka, Luitz et~al.}}]{blaha2001}
\bibinfo{author}{\bibfnamefont{P.}~\bibnamefont{Blaha}},
  \bibinfo{author}{\bibfnamefont{K.}~\bibnamefont{Schwarz}},
  \bibinfo{author}{\bibfnamefont{G.~K.} \bibnamefont{Madsen}},
  \bibinfo{author}{\bibfnamefont{D.}~\bibnamefont{Kvasnicka}},
  \bibinfo{author}{\bibfnamefont{J.}~\bibnamefont{Luitz}},
  \bibnamefont{et~al.}, \bibinfo{journal}{An augmented plane wave+ local
  orbitals program for calculating crystal properties}  (\bibinfo{year}{2001}).

\bibitem[{\citenamefont{Perdew et~al.}(1996)\citenamefont{Perdew, Burke, and
  Ernzerhof}}]{perdew1996}
\bibinfo{author}{\bibfnamefont{J.~P.} \bibnamefont{Perdew}},
  \bibinfo{author}{\bibfnamefont{K.}~\bibnamefont{Burke}}, \bibnamefont{and}
  \bibinfo{author}{\bibfnamefont{M.}~\bibnamefont{Ernzerhof}},
  \bibinfo{journal}{Physical review letters} \textbf{\bibinfo{volume}{77}},
  \bibinfo{pages}{3865} (\bibinfo{year}{1996}).

\bibitem[{\citenamefont{Sala et~al.}(2021)\citenamefont{Sala, Stone, Rai, May,
  Laurell, Garlea, Butch, Lumsden, Ehlers, Pokharel et~al.}}]{Sala2021}
\bibinfo{author}{\bibfnamefont{G.}~\bibnamefont{Sala}},
  \bibinfo{author}{\bibfnamefont{M.~B.} \bibnamefont{Stone}},
  \bibinfo{author}{\bibfnamefont{B.~K.} \bibnamefont{Rai}},
  \bibinfo{author}{\bibfnamefont{A.~F.} \bibnamefont{May}},
  \bibinfo{author}{\bibfnamefont{P.}~\bibnamefont{Laurell}},
  \bibinfo{author}{\bibfnamefont{V.~O.} \bibnamefont{Garlea}},
  \bibinfo{author}{\bibfnamefont{N.~P.} \bibnamefont{Butch}},
  \bibinfo{author}{\bibfnamefont{M.~D.} \bibnamefont{Lumsden}},
  \bibinfo{author}{\bibfnamefont{G.}~\bibnamefont{Ehlers}},
  \bibinfo{author}{\bibfnamefont{G.}~\bibnamefont{Pokharel}},
  \bibnamefont{et~al.}, \bibinfo{journal}{Nature Communications}
  \textbf{\bibinfo{volume}{12}}, \bibinfo{pages}{171} (\bibinfo{year}{2021}),
  ISSN \bibinfo{issn}{2041-1723},
  \urlprefix\url{https://doi.org/10.1038/s41467-020-20335-5}.

\bibitem[{\citenamefont{Pokharel et~al.}(2020)\citenamefont{Pokharel,
  Arachchige, Williams, May, Fishman, Sala, Calder, Ehlers, Parker, Hong
  et~al.}}]{Pokharel2020PRL}
\bibinfo{author}{\bibfnamefont{G.}~\bibnamefont{Pokharel}},
  \bibinfo{author}{\bibfnamefont{H.~S.} \bibnamefont{Arachchige}},
  \bibinfo{author}{\bibfnamefont{T.~J.} \bibnamefont{Williams}},
  \bibinfo{author}{\bibfnamefont{A.~F.} \bibnamefont{May}},
  \bibinfo{author}{\bibfnamefont{R.~S.} \bibnamefont{Fishman}},
  \bibinfo{author}{\bibfnamefont{G.}~\bibnamefont{Sala}},
  \bibinfo{author}{\bibfnamefont{S.}~\bibnamefont{Calder}},
  \bibinfo{author}{\bibfnamefont{G.}~\bibnamefont{Ehlers}},
  \bibinfo{author}{\bibfnamefont{D.~S.} \bibnamefont{Parker}},
  \bibinfo{author}{\bibfnamefont{T.}~\bibnamefont{Hong}}, \bibnamefont{et~al.},
  \bibinfo{journal}{Phys. Rev. Lett.} \textbf{\bibinfo{volume}{125}},
  \bibinfo{pages}{167201} (\bibinfo{year}{2020}),
  \urlprefix\url{https://link.aps.org/doi/10.1103/PhysRevLett.125.167201}.

\bibitem[{\citenamefont{Pokharel et~al.}(2018)\citenamefont{Pokharel, May,
  Parker, Calder, Ehlers, Huq, Kimber, Arachchige, Poudel, McGuire
  et~al.}}]{Pokharel2018PRB}
\bibinfo{author}{\bibfnamefont{G.}~\bibnamefont{Pokharel}},
  \bibinfo{author}{\bibfnamefont{A.~F.} \bibnamefont{May}},
  \bibinfo{author}{\bibfnamefont{D.~S.} \bibnamefont{Parker}},
  \bibinfo{author}{\bibfnamefont{S.}~\bibnamefont{Calder}},
  \bibinfo{author}{\bibfnamefont{G.}~\bibnamefont{Ehlers}},
  \bibinfo{author}{\bibfnamefont{A.}~\bibnamefont{Huq}},
  \bibinfo{author}{\bibfnamefont{S.~A.~J.} \bibnamefont{Kimber}},
  \bibinfo{author}{\bibfnamefont{H.~S.} \bibnamefont{Arachchige}},
  \bibinfo{author}{\bibfnamefont{L.}~\bibnamefont{Poudel}},
  \bibinfo{author}{\bibfnamefont{M.~A.} \bibnamefont{McGuire}},
  \bibnamefont{et~al.}, \bibinfo{journal}{Phys. Rev. B}
  \textbf{\bibinfo{volume}{97}}, \bibinfo{pages}{134117}
  (\bibinfo{year}{2018}),
  \urlprefix\url{https://link.aps.org/doi/10.1103/PhysRevB.97.134117}.

\bibitem[{\citenamefont{Yan et~al.}(2020)\citenamefont{Yan, Liu, Parker, Wu,
  Aczel, Matsuda, McGuire, and Sales}}]{Yan2020PRM}
\bibinfo{author}{\bibfnamefont{J.-Q.} \bibnamefont{Yan}},
  \bibinfo{author}{\bibfnamefont{Y.~H.} \bibnamefont{Liu}},
  \bibinfo{author}{\bibfnamefont{D.~S.} \bibnamefont{Parker}},
  \bibinfo{author}{\bibfnamefont{Y.}~\bibnamefont{Wu}},
  \bibinfo{author}{\bibfnamefont{A.~A.} \bibnamefont{Aczel}},
  \bibinfo{author}{\bibfnamefont{M.}~\bibnamefont{Matsuda}},
  \bibinfo{author}{\bibfnamefont{M.~A.} \bibnamefont{McGuire}},
  \bibnamefont{and} \bibinfo{author}{\bibfnamefont{B.~C.} \bibnamefont{Sales}},
  \bibinfo{journal}{Phys. Rev. Materials} \textbf{\bibinfo{volume}{4}},
  \bibinfo{pages}{054202} (\bibinfo{year}{2020}),
  \urlprefix\url{https://link.aps.org/doi/10.1103/PhysRevMaterials.4.054202}.

\bibitem[{\citenamefont{Moreau et~al.}(1975)\citenamefont{Moreau, Parth{\'{e}},
  and Paccard}}]{Moreau1975ACB}
\bibinfo{author}{\bibfnamefont{J.~M.} \bibnamefont{Moreau}},
  \bibinfo{author}{\bibfnamefont{E.}~\bibnamefont{Parth{\'{e}}}},
  \bibnamefont{and} \bibinfo{author}{\bibfnamefont{D.}~\bibnamefont{Paccard}},
  \bibinfo{journal}{Acta Crystallographica Section B}
  \textbf{\bibinfo{volume}{31}}, \bibinfo{pages}{747} (\bibinfo{year}{1975}).

\bibitem[{\citenamefont{Sichevich}(1984)}]{Sichevich1984Dopovidi}
\bibinfo{author}{\bibfnamefont{O.~M.} \bibnamefont{Sichevich}},
  \bibinfo{journal}{Dopov. Akad. Nauk Ukr. RSR, Ser. B}
  \textbf{\bibinfo{volume}{12}} (\bibinfo{year}{1984}).

\bibitem[{\citenamefont{Johnston}({2012})}]{Johnston2012PRL}
\bibinfo{author}{\bibfnamefont{D.~C.} \bibnamefont{Johnston}},
  \bibinfo{journal}{{Phys. Rev. Lett.}} \textbf{\bibinfo{volume}{{109}}},
  \bibinfo{pages}{077201} (\bibinfo{year}{{2012}}), ISSN
  \bibinfo{issn}{{0031-9007}}.

\bibitem[{\citenamefont{Blundell}(2001)}]{Blundell2001Magnetism}
\bibinfo{author}{\bibfnamefont{S.~J.} \bibnamefont{Blundell}},
  \emph{\bibinfo{title}{Magnetism in Condensed-Matter}}
  (\bibinfo{publisher}{Oxford}, \bibinfo{year}{2001}).

\bibitem[{\citenamefont{Cheng et~al.}(2012)\citenamefont{Cheng, Hu, Yuan, Dong,
  Fang, Chen, Xu, Shi, Zheng, Luo et~al.}}]{Cheng2012PRB}
\bibinfo{author}{\bibfnamefont{B.}~\bibnamefont{Cheng}},
  \bibinfo{author}{\bibfnamefont{B.~F.} \bibnamefont{Hu}},
  \bibinfo{author}{\bibfnamefont{R.~H.} \bibnamefont{Yuan}},
  \bibinfo{author}{\bibfnamefont{T.}~\bibnamefont{Dong}},
  \bibinfo{author}{\bibfnamefont{A.~F.} \bibnamefont{Fang}},
  \bibinfo{author}{\bibfnamefont{Z.~G.} \bibnamefont{Chen}},
  \bibinfo{author}{\bibfnamefont{G.}~\bibnamefont{Xu}},
  \bibinfo{author}{\bibfnamefont{Y.~G.} \bibnamefont{Shi}},
  \bibinfo{author}{\bibfnamefont{P.}~\bibnamefont{Zheng}},
  \bibinfo{author}{\bibfnamefont{J.~L.} \bibnamefont{Luo}},
  \bibnamefont{et~al.}, \bibinfo{journal}{Phys. Rev. B}
  \textbf{\bibinfo{volume}{85}}, \bibinfo{pages}{144426}
  (\bibinfo{year}{2012}),
  \urlprefix\url{https://link.aps.org/doi/10.1103/PhysRevB.85.144426}.

\bibitem[{\citenamefont{Barton et~al.}(2013)\citenamefont{Barton, Seshadri,
  Llobet, and Suchomel}}]{Barton2013PRB}
\bibinfo{author}{\bibfnamefont{P.~T.} \bibnamefont{Barton}},
  \bibinfo{author}{\bibfnamefont{R.}~\bibnamefont{Seshadri}},
  \bibinfo{author}{\bibfnamefont{A.}~\bibnamefont{Llobet}}, \bibnamefont{and}
  \bibinfo{author}{\bibfnamefont{M.~R.} \bibnamefont{Suchomel}},
  \bibinfo{journal}{Phys. Rev. B} \textbf{\bibinfo{volume}{88}},
  \bibinfo{pages}{024403} (\bibinfo{year}{2013}),
  \urlprefix\url{https://link.aps.org/doi/10.1103/PhysRevB.88.024403}.

\bibitem[{\citenamefont{Bartashevich et~al.}(1983)\citenamefont{Bartashevich,
  Deryagin, Kudrevatykh, and Tarasov}}]{bartashevich1983high}
\bibinfo{author}{\bibfnamefont{M.}~\bibnamefont{Bartashevich}},
  \bibinfo{author}{\bibfnamefont{A.}~\bibnamefont{Deryagin}},
  \bibinfo{author}{\bibfnamefont{N.}~\bibnamefont{Kudrevatykh}},
  \bibnamefont{and} \bibinfo{author}{\bibfnamefont{E.}~\bibnamefont{Tarasov}},
  \bibinfo{journal}{Sov. Phys.-JETP} \textbf{\bibinfo{volume}{57}},
  \bibinfo{pages}{662} (\bibinfo{year}{1983}).

\bibitem[{\citenamefont{Goto et~al.}(1992)\citenamefont{Goto, Nakao,
  Sakakibara, Ito, and Yamada}}]{goto1992field}
\bibinfo{author}{\bibfnamefont{T.}~\bibnamefont{Goto}},
  \bibinfo{author}{\bibfnamefont{K.}~\bibnamefont{Nakao}},
  \bibinfo{author}{\bibfnamefont{T.}~\bibnamefont{Sakakibara}},
  \bibinfo{author}{\bibfnamefont{M.}~\bibnamefont{Ito}}, \bibnamefont{and}
  \bibinfo{author}{\bibfnamefont{I.}~\bibnamefont{Yamada}},
  \bibinfo{journal}{Physica B: Condensed Matter}
  \textbf{\bibinfo{volume}{177}}, \bibinfo{pages}{373} (\bibinfo{year}{1992}).

\bibitem[{\citenamefont{Bartashevich et~al.}(1994)\citenamefont{Bartashevich,
  Katori, Goto, Yamamoto, and Yamaguchi}}]{bartashevich1994ultrahigh}
\bibinfo{author}{\bibfnamefont{M.}~\bibnamefont{Bartashevich}},
  \bibinfo{author}{\bibfnamefont{H.~A.} \bibnamefont{Katori}},
  \bibinfo{author}{\bibfnamefont{T.}~\bibnamefont{Goto}},
  \bibinfo{author}{\bibfnamefont{I.}~\bibnamefont{Yamamoto}}, \bibnamefont{and}
  \bibinfo{author}{\bibfnamefont{M.}~\bibnamefont{Yamaguchi}},
  \bibinfo{journal}{Physica B: Condensed Matter}
  \textbf{\bibinfo{volume}{201}}, \bibinfo{pages}{135} (\bibinfo{year}{1994}).

\bibitem[{\citenamefont{Wipf}(2012)}]{wipf2012statistical}
\bibinfo{author}{\bibfnamefont{A.}~\bibnamefont{Wipf}},
  \emph{\bibinfo{title}{Statistical approach to quantum field theory: an
  introduction}}, vol. \bibinfo{volume}{864} (\bibinfo{publisher}{Springer},
  \bibinfo{year}{2012}).

\bibitem[{\citenamefont{Kittel et~al.}(1996)\citenamefont{Kittel, McEuen, and
  McEuen}}]{kittel1996introduction}
\bibinfo{author}{\bibfnamefont{C.}~\bibnamefont{Kittel}},
  \bibinfo{author}{\bibfnamefont{P.}~\bibnamefont{McEuen}}, \bibnamefont{and}
  \bibinfo{author}{\bibfnamefont{P.}~\bibnamefont{McEuen}},
  \emph{\bibinfo{title}{Introduction to solid state physics}},
  vol.~\bibinfo{volume}{8} (\bibinfo{publisher}{Wiley New York},
  \bibinfo{year}{1996}).

\bibitem[{\citenamefont{Lamichhane et~al.}(2018)\citenamefont{Lamichhane,
  Taufour, Palasyuk, Lin, Bud'ko, and Canfield}}]{Lamichhane2018PRA}
\bibinfo{author}{\bibfnamefont{T.~N.} \bibnamefont{Lamichhane}},
  \bibinfo{author}{\bibfnamefont{V.}~\bibnamefont{Taufour}},
  \bibinfo{author}{\bibfnamefont{A.}~\bibnamefont{Palasyuk}},
  \bibinfo{author}{\bibfnamefont{Q.}~\bibnamefont{Lin}},
  \bibinfo{author}{\bibfnamefont{S.~L.} \bibnamefont{Bud'ko}},
  \bibnamefont{and} \bibinfo{author}{\bibfnamefont{P.~C.}
  \bibnamefont{Canfield}}, \bibinfo{journal}{Phys. Rev. Applied}
  \textbf{\bibinfo{volume}{9}}, \bibinfo{pages}{024023} (\bibinfo{year}{2018}),
  \urlprefix\url{https://link.aps.org/doi/10.1103/PhysRevApplied.9.024023}.

\bibitem[{\citenamefont{Lamichhane et~al.}(2020)\citenamefont{Lamichhane,
  Taufour, Palasyuk, Bud'ko, and Canfield}}]{Lamichhane2020PM}
\bibinfo{author}{\bibfnamefont{T.~N.} \bibnamefont{Lamichhane}},
  \bibinfo{author}{\bibfnamefont{V.}~\bibnamefont{Taufour}},
  \bibinfo{author}{\bibfnamefont{A.}~\bibnamefont{Palasyuk}},
  \bibinfo{author}{\bibfnamefont{S.~L.} \bibnamefont{Bud'ko}},
  \bibnamefont{and} \bibinfo{author}{\bibfnamefont{P.~C.}
  \bibnamefont{Canfield}}, \bibinfo{journal}{Philosophical Magazine}
  \textbf{\bibinfo{volume}{100}}, \bibinfo{pages}{1607} (\bibinfo{year}{2020}),
  ISSN \bibinfo{issn}{1478-6435},
  \urlprefix\url{https://doi.org/10.1080/14786435.2020.1727973}.

\bibitem[{\citenamefont{Pandey and Parker}(2018)}]{Pandey2018PRA}
\bibinfo{author}{\bibfnamefont{T.}~\bibnamefont{Pandey}} \bibnamefont{and}
  \bibinfo{author}{\bibfnamefont{D.~S.} \bibnamefont{Parker}},
  \bibinfo{journal}{Phys. Rev. Applied} \textbf{\bibinfo{volume}{10}},
  \bibinfo{pages}{034038} (\bibinfo{year}{2018}),
  \urlprefix\url{https://link.aps.org/doi/10.1103/PhysRevApplied.10.034038}.

\bibitem[{\citenamefont{Moon et~al.}(2010)\citenamefont{Moon, Shin, Parker,
  Choi, Mazin, Lee, Kim, Sung, Cho, Khim et~al.}}]{Moon2010PRB}
\bibinfo{author}{\bibfnamefont{S.~J.} \bibnamefont{Moon}},
  \bibinfo{author}{\bibfnamefont{J.~H.} \bibnamefont{Shin}},
  \bibinfo{author}{\bibfnamefont{D.}~\bibnamefont{Parker}},
  \bibinfo{author}{\bibfnamefont{W.~S.} \bibnamefont{Choi}},
  \bibinfo{author}{\bibfnamefont{I.~I.} \bibnamefont{Mazin}},
  \bibinfo{author}{\bibfnamefont{Y.~S.} \bibnamefont{Lee}},
  \bibinfo{author}{\bibfnamefont{J.~Y.} \bibnamefont{Kim}},
  \bibinfo{author}{\bibfnamefont{N.~H.} \bibnamefont{Sung}},
  \bibinfo{author}{\bibfnamefont{B.~K.} \bibnamefont{Cho}},
  \bibinfo{author}{\bibfnamefont{S.~H.} \bibnamefont{Khim}},
  \bibnamefont{et~al.}, \bibinfo{journal}{Phys. Rev. B}
  \textbf{\bibinfo{volume}{81}}, \bibinfo{pages}{205114}
  (\bibinfo{year}{2010}),
  \urlprefix\url{https://link.aps.org/doi/10.1103/PhysRevB.81.205114}.

\end{thebibliography}

\end{document}